\documentclass[a4,11pt]{cip-submit}  
\usepackage{color,soul}
\usepackage{afterpage}

\usepackage{wrapfig}
\usepackage{wasysym}
\usepackage{enumitem}
\usepackage{adjustbox}
\usepackage{ragged2e}
\usepackage[svgnames,table]{xcolor}
\usepackage{tikz}
\usepackage{longtable}
\usepackage{changepage}
\usepackage{setspace}
\usepackage{hhline}
\usepackage{multicol}
\usepackage{tabto}
\usepackage{float}
\usepackage{multirow}
\usepackage{makecell}
\usepackage{tabularx} 
\usepackage{upgreek} 
\usepackage{mathptmx}
\DeclareSymbolFont{epsilon}{OML}{cmm}{m}{it}
\DeclareMathSymbol{\epsilon}{\mathord}{epsilon}{"0F}
\def\DD{D\kern-.7em\raise0.25ex\hbox{\char '55}\kern.33em}
\usepackage{hyperref}
\hypersetup{
	colorlinks=false,
	pdfborder={0 0 0}
}
\usepackage{graphicx,subfigure,wrapfig,epstopdf}
\def\Mr{\uppercase}
\usepackage{graphics}
\usepackage{wasysym}
\usepackage{amsmath,amsxtra,amssymb,latexsym,amscd,color,cite,bm}
\usepackage[mathscr]{eucal}
\usepackage{array}
\usepackage{multirow}
\usepackage{cite} 

\def\doi#1{\href{http://dx.doi.org/#1}{doi:#1}}
\def\titles#1{\title{\large\bf\noindent #1}}
\def\authors#1{\author{\begin{flushleft}{#1}\end{flushleft}}}
\def\authord#1#2{\indent\Mr{#1}$^{#2}$}
\def\addressed#1#2{\\[1mm]\textit{$\!\!\!^{#1}$\indent#2}}

\def\Email{$^{\dagger}$}
\def\email#1{\bigskip\href{mailto:#1}{\!\!\Email\textit{E-mail:}~{#1}}\\[3mm]}
 
\def\PublicationInformation#1#2#3{\\[3mm]\textit{\indent Received~#1}\\[1mm]
	\textit{Accepted for publication~#2}\\[1mm]
	\textit{Published~#3}}

\def\Keywords#1{\\[.2cm] \textnormal{Keywords:~{#1}}.} 

\def\and{$\text{\tiny AND }$}

\newcommand{\sout}[1]{\unskip}

\def\ed{
	\bibliographystyle{cip-sty-2019}
	\bibliography{references-database-name}
\end{document}
}
\def\abstractbreak{\break
\end{abstract} 
\newpage
}
\def\CorrEmail#1{\\[4mm]
\textit{E-mail:}~$^\dag${#1}}

\begin{document}
	\Year{2023}
	\Page{1}\Endpage{4}
	\titles{Direct triple-$\alpha$ process in non-adiabatic approach}
	\authors{
	\authord{M. Katsuma}{1,2}\Email
	\newline
	\addressed{1}{Institut d'Astronomie et d'Astrophysique, Universit\'e Libre de Bruxelles, B1050 Brussels, Belgium}
	\addressed{2}{Advanced Mathematical Institute, Osaka City University, Osaka 558-8585, Japan}
\CorrEmail{mkatsuma@gmail.com}
\PublicationInformation{March 10, 2023}{}{}
	}
	\maketitle
	\markboth{Direct triple-$\alpha$ process in non-adiabatic approach}{M. Katsuma}

\begin{abstract}
  Triple-$\alpha$ reaction rates have been determined well with the sequential process via the narrow resonances.
  However, the direct triple-$\alpha$ process at off-resonant energies still remains in unsolved problems.
  In the present report, the direct triple-$\alpha$ contribution is estimated with a non-adiabatic method, and it is found to be 10$^{-15}$--10$^{-3}$ pb order in photodisintegration cross sections of $^{12}$C(2$^+_1 \rightarrow$ 0$^+$) for $0.15 < E < 0.35$ MeV.
  This is far below the values predicted by the recent adiabatic models.
  In spite of the large difference, the derived rates are found to be concordant with NACRE at the helium burning temperatures.
\end{abstract}

\section{\Mr{Introduction}}
  The triple-$\alpha$ reaction plays an important role in nucleosynthesis heavier than $^{12}$C, because no stable nuclei exist in mass number $A=5$ and $A=8$ \cite{Hoy54}.
  This reaction, followed by $^{12}$C($\alpha$,$\gamma$)$^{16}$O \cite{Kat08}, controls C/O ratio at the end of helium burning phase in stars, and it affects up to the nucleosynthesis in supernova explosion.
  In contrast to $^{12}$C($\alpha$,$\gamma$)$^{16}$O, the triple-$\alpha$ reaction is currently well-understood through the experimental studies of the 0$^+_2$ state in $^{12}$C ($E_r=0.379$ MeV) ({\it e.g.} \cite{Del17,Nacre}),
  {\it i.e.}, the reaction rates have been determined well with the sequential process via the narrow resonances: $\alpha+\alpha\rightarrow ^8$Be(0$^+_1$), $\alpha$+$^8$Be$\rightarrow ^{12}$C(0$^+_2$).
  Pioneering works of the reaction rates have been performed by \cite{CF88,Nom85}, and their experimental upgrade has been given by NACRE \cite{Nacre}.

  Apart from the sequential process, the triple-$\alpha$ reaction from 3$\alpha$ continuum states is referred to as the direct triple-$\alpha$ process: $\alpha+\alpha+\alpha\rightarrow ^{12}$C.
  This process is generally expected to be very slow, because three $\alpha$-particles almost simultaneously collide and fuse into a $^{12}$C nucleus.
  So, this process is neglected or is treated in some approximations.
  For the theoretical studies, formulae in hyper-spherical coordinates have been applied to tackle the 3$\alpha$ continuum problem ({\it e.g.} \cite{Fed96}), and their adiabatic approaches have paved the way for a non-adiabatic approach of \cite{Ngu13}.
  The Coulomb modified Faddeev (CMF) \cite{Ish14} and adiabatic channel function (ACF) expansion method \cite{Sun16} may have also achieved the successful progress quantitatively.
  However, non-adiabatic quantum-mechanical description at off-resonant energies still seem to remain in unsolved problems.
  In the present report, the contribution of the direct 3$\alpha$ process is estimated with a non-adiabatic Faddeev hyper-spherical harmonics and $R$-matrix (HHR$^\ast$) expansion method \cite{Tho09,Des}.
  At the same time, the difference between the non-adiabatic and adiabatic calculations is discussed.

\section{\Mr{Faddeev HHR$^\ast$ expansion method}}
  Before discussing the calculated results, let me describe HHR$^\ast$, briefly.
  The triple-$\alpha$ system satisfies the three-body Schr\"odinger equation, $(H_{3\alpha} -E)\Psi_{lm} = 0$.
  $H_{3\alpha}$ is three-body Hamiltonian; $E$ is the center-of-mass energy to the 3$\alpha$ threshold in $^{12}$C; $l$ is spin of the states in $^{12}$C; $m$ is the projection of $l$.
  This equation is expressed as the so-called Faddeev equations, consisting of three components.
  Due to the symmetric 3$\alpha$ system, three identical sets of equations are found, and they are expressed in a similar form of ordinary coupled-channel (CC) equations for inelastic scattering, ({\it e.g.} \cite{Kat02,Sat}), after translating Jacobi coordinates (${\bf{x}}_3, {\bf{y}}_3$) into hyper-spherical coordinates ($\rho, \Omega_5$), $\Omega_5 \equiv (\theta_3, \hat{\bf{x}}_3, \hat{\bf{y}}_3)$.
  The deduced CC equations of the hyper-radial waves, $\chi^l_{\gamma}(\rho)$, are given in
  \begin{eqnarray}
    \left[\, T_\gamma +U^l_{\gamma \gamma}(\rho)-\epsilon \,\right] \chi^l_{\gamma}(\rho) &=&
    -\mathop{\sum}_{\gamma^\prime \ne\gamma} U^l_{\gamma^\prime \gamma}(\rho) \chi^l_{\gamma^\prime}(\rho),
    \label{eq:hscc} 
  \end{eqnarray}
  where $T_\gamma=d^2/d\rho^2 -(K+3/2)(K+5/2)/\rho^2$; $U^l_{\gamma^\prime \gamma}$ are the coupling potentials; $\epsilon = -2m_NE/\hbar^2$; $m_N$ is nucleon mass; $K$ is hyper-angular momentum; $\gamma$ is a label of channels.
  $U^l_{\gamma^\prime \gamma}$ are calculated from $\alpha+\alpha$ and 3$\alpha$ potentials, and they are the same as \cite{Ngu13} except that the strength of 3$\alpha$ potentials are $-20.145$ MeV ($-19.46$ MeV \cite{Ngu13}) for 0$^+$ and $-16.36$ MeV ($-15.94$ MeV \cite{Ngu13}) for 2$^+$.
  Using the hyper-harmonic function $\Phi^\gamma_{lm}(\Omega_5)$ \cite{Tho09,Des}, the basis functions are defined by
  \begin{eqnarray}
    \Psi_{lm} &=& \rho^{-5/2}\mathop{\sum}_{\gamma n} c^n_\gamma \varphi^K_n(\rho) \Phi^\gamma_{lm}(\Omega_5),
    \label{eq:hhex} 
  \end{eqnarray}
  where $\varphi^K_n(\rho)$ are harmonic oscillator wavefunctions in hyper-spherical coordinates, $\chi^l_{\gamma} = \mathop{\sum}_n c^n_\gamma \varphi^K_n$.
  The results are independent of $\varphi^K_n(\rho)$, if a large number of basis are used.
  The CC equations~\eqref{eq:hscc}, briefly rewritten as $(\bf{T}+\bf{U})\bf{X} = \epsilon \bf{X}$, are solved by the matrix diagonalization.
  The matrix size of the present calculation is (8,800$\times$8,800) for 0$^+$ in $^{12}$C.
  In the $R$-matrix expansion method, the continuum states with scattering boundary condition are expanded by the resultant eigenfunctions,
  \begin{eqnarray}
    \chi^{l\hspace{1mm}in}_{\alpha\beta}(k,\rho) &=& \mathop{\sum}_i A_{i \beta}(k) \chi^l_{\alpha i}(\rho),
    \label{eq:chi_in} 
  \end{eqnarray}
  \vspace{-3mm}
  \begin{equation}
    A_{i \beta}(k) = \frac{\hbar^2}{2m_N} \frac{1}{E(l^+_i)-E} \mathop{\sum}_\gamma \chi^l_{\gamma i}(a_c)
    \Big[\,H^{-\hspace{1mm}\prime}_{K+3/2}(\eta_\gamma; ka_c) \delta_{\gamma\beta} -S^l_{\gamma\beta}(E,a_c)
      H^{+\hspace{1mm}\prime}_{K+3/2}(\eta_\gamma; ka_c)\,\Big],
    \label{eq:aib} 
  \end{equation}
  where $\chi^l_{\alpha i}(\rho)$ and $E(l^+_i)$ are the eigenfunctions and eigen-energy, respectively.
  $\alpha$ and $\beta$ are the channel labels.
  $H^{\pm}$ are the incoming ($-$) and outgoing ($+$) Coulomb wavefunctions; $\eta_\gamma$ is the Sommerfeld parameter; $k$ is the hyper-momentum, $k=(2m_N E/\hbar^2)^{1/2}$; $a_c$ is the channel radius.
  $S^l_{\gamma\beta}(E,a_c)$ is the $S$-matrix, defined by the derived $R$-matrix at $\rho=a_c$.

  To include the long-range Coulomb couplings, the CC equations in the external region are solved numerically from $\rho = a_c$ to $\rho_m$ by using the $R$-matrix propagation technique \cite{Ngu13}.
  The external wavefunctions are expanded by the resulting linearly-independent solutions $\chi^{l\hspace{1mm}ext}_{\gamma\gamma^\prime} (k,\rho)$, and the coefficients of expansion $C_{\gamma^\prime\gamma_0}(k)$ are obtained by matching to the asymptotic form,
  \begin{eqnarray}
    \tilde{\chi}^l_{\gamma\gamma_0}(k,\rho) &\rightarrow& \frac{i}{2}
    \Big[\,I^{(\gamma_0)}_{\gamma, K+3/2}(\eta_\gamma; k\rho_m) -\mathop{\sum}_{\gamma^\prime} S^l_{\gamma^\prime \gamma_0}(E)
      O^{(\gamma^\prime)}_{\gamma, K+3/2}(\eta_\gamma; k\rho_m)\,\Big],
    \label{eq:asymptotic} 
  \end{eqnarray}
  where $O$ and $I=O^\ast$ are the coupled-Coulomb waves \cite{Sat}, $O^{(\gamma^\prime)}_\gamma = a_\gamma^{(\gamma^\prime)}(k,\rho) H^{+}_\gamma$.
  In the present report, the global back propagation \cite{Ngu13} from $\rho=\rho_m$ to $a_c$ is not used.
  If the screening potential \cite{Ngu13} is adopted to reduce the strength of Coulomb couplings at large $\rho$, $a_\gamma^{(\gamma^\prime)}(k,\rho_m) =\delta_{\gamma\gamma^\prime}$ can be used.
  This is effective if the off-diagonal part of coupling potentials is relatively small at $\rho_m$, compared with $E$, ({\it e.g.}~\cite{Kat02}).
  Multiplying eq.~\eqref{eq:chi_in} by $C_{\gamma^\prime\gamma_0}(k)$, I obtain the interior scattering waves including the long-range Coulomb couplings, $\tilde{\chi}^{l\hspace{1mm} in}_{\gamma\gamma_0}(k,\rho) = \mathop{\sum}_{\gamma^\prime}C_{\gamma^\prime\gamma_0}(k) \chi^{l\hspace{1mm}in}_{\gamma\gamma^\prime}(k,\rho)$.

  The photodisintegration cross sections of $^{12}$C(2$^+_1\rightarrow 0^+$) are calculated from
  \begin{eqnarray}
    \sigma_g(E) &=& \frac{2\pi^2}{75}\left(\frac{E_g}{\hbar c}\right)^3
    \frac{e^2}{\hbar v} \mathop{\sum}_{\gamma_0}
    \Big| \mathop{\sum}_{\gamma^\prime\gamma} \big[ M^{2^+_1\,0^+\,{Int}}_{\gamma^\prime\gamma\gamma_0}(k) + M^{2^+_1\,0^+\,{Ext}}_{\gamma^\prime\gamma\gamma_0}(k) \big]
    F_{\gamma^\prime\gamma} \Big|^2,
    \label{eq:sig} 
  \end{eqnarray}
  where $E_g$ is the $\gamma$-ray energy, $E_g=E-E(2^+_1)$.
  $c$ is speed of light, and $v=(2E/m_N)^{1/2}$.
  $F_{\gamma^\prime\gamma}$ are the factors stemming from the hyper-angle part.
  $M^{2^+_1\,0^+\,{Int}}_{\gamma^\prime\gamma\gamma_0}$ and $M^{2^+_1\,0^+\,{Ext}}_{\gamma^\prime\gamma\gamma_0}$ are the internal and external components of the hyper-radial part.
  To execute stable calculations, quadruple precision is required.
  The energy-averaged reaction rates $\langle R_{3\alpha} \rangle$ are calculated from the resultant $\sigma_g$.

  \begin{figure}[t]
    \minipage{0.485\textwidth}
    \centering
    \includegraphics[width=0.89\linewidth]{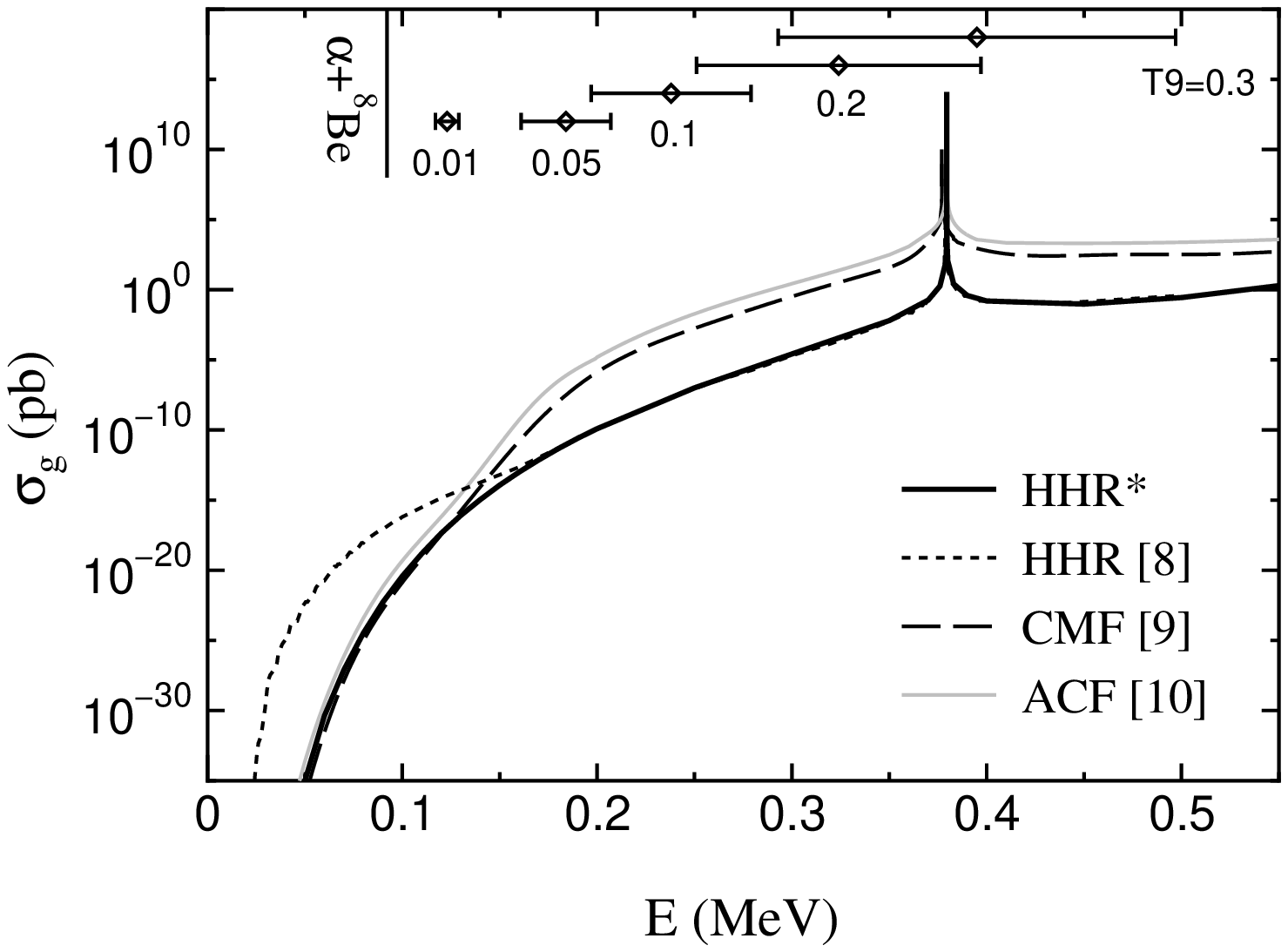}
    \caption{\label{fig:sig}
      Photodisintegration cross sections of $^{12}$C (2$^+_1 \rightarrow$ 0$^+$).
      The solid curve is the result obtained from HHR$^\ast$.
      The dotted, dashed and gray curves are from \cite{Ngu13,Ish14,Sun16}.
    }
    \endminipage
    \hfill
    \quad
    \minipage{0.485\textwidth}
    \centering
    \includegraphics[width=0.8\linewidth]{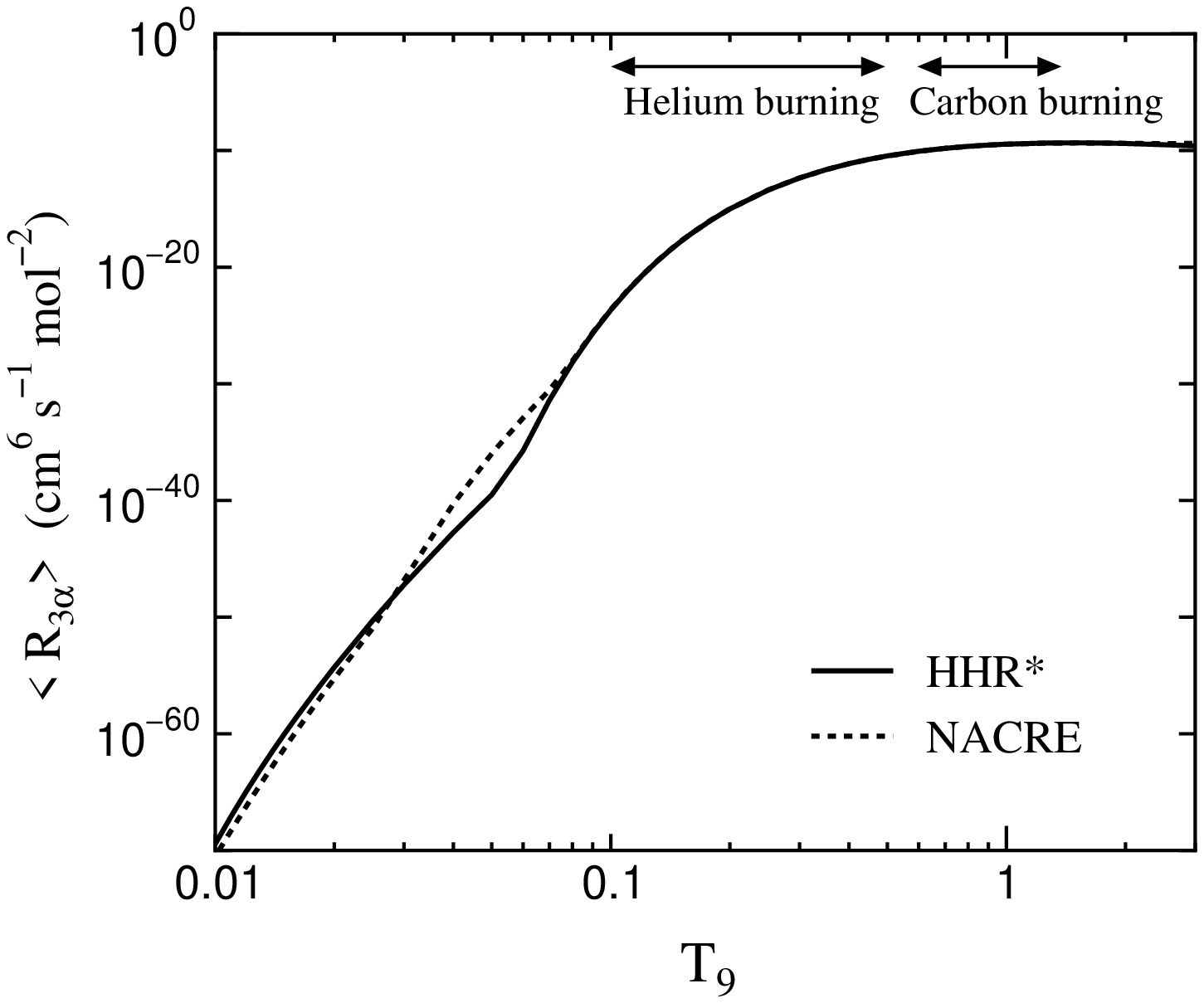}
    \caption{\label{fig:rate}
      Triple-$\alpha$ reaction rates.
      The solid curve is the result obtained from HHR$^\ast$.
      The NACRE rates \cite{Nacre} are shown by the dotted curve.
    }
    \endminipage
    \hfill
  \end{figure}

\section{\Mr{Results}}
  The calculated photodisintegration of $^{12}$C(2$^+_1 \rightarrow 0^+$) is shown by the solid curve in Fig.~\ref{fig:sig}.
  The prominent narrow resonance of 0$^+_2$ is found at $E=0.3795$ MeV, and the smoothly varying non-resonant cross sections are obtained at off-resonant energies.
  The calculated $\alpha$- and $\gamma$-widths are $\Gamma_\alpha=5.1$ eV and $\Gamma_\gamma=4.3$ meV, comparable to the experimental data \cite{Nacre}: $\Gamma_\alpha=8.3$ eV and $\Gamma_\gamma=3.7$ meV.
  For $0.15 < E < 0.35$ MeV, I find $\sigma_g = 10^{-15}$--$10^{-3}$ pb order of cross sections.
  This result is almost identical to the dotted curve of \cite{Ngu13}.
  The result below $E=0.15$ MeV seems similar to CMF (dashed curve) and ACF (gray curve).
  On the other hand, the present result above $E=0.15$ MeV is much smaller than the values predicted by CMF and ACF.
  CMF has been developed below the three-body threshold, {\it e.g.} low-energy p+d reactions.
  The internal motion of $^8$Be in break-up channels is assumed to be localized within a certain range.
  ACF has the feature of $\alpha$+$^8$Be for $\rho < 150$ fm and 3$\alpha$ for large $\rho$, and the resonance and bound states are expanded with the adiabatic basis.
  Judging from their theoretical approaches, most of the differences above $E=0.15$ MeV seem to stem from the internal adiabatic feature.
  The adiabatic approach to $^8$Be continuum states makes the assumed long resonant tail of 0$^+_2$, leading to the sequential decay process at off-resonant energies, and it might have enhanced the photodisintegration cross sections unexpectedly.
  The enhancement of \cite{Ngu13} for $E<0.15$ MeV seems to be caused by the redundant propagation.

  In spite of the large difference, the derived rates are found to be consistent with NACRE for $0.08 < T_9 < 3$, including the helium burning temperatures (see Fig.~\ref{fig:rate}).
  In contrast, the present result is reduced by 10$^{-4}$ at $T_9 = 0.05$, because $\sigma_g$ are reduced from CMF and ACF with the sequential process at $E = 0.18$ MeV.
  Due to the strong influence of 0$^+_2$, the difference in $\sigma_g$ for $E > 0.2$ MeV cannot be found in the rates.

\section{\Mr{Conclusion}}
  From the present calculation, I have found that the direct triple-$\alpha$ contribution is 10$^{-15}$--10$^{-3}$ pb order in the photodisintegration cross sections of $^{12}$C (2$^+_1 \rightarrow 0^+$) for $0.15 < E < 0.35$ MeV.
  This is far below the values predicted by CMF and ACF that include the assumed long resonant tail of 0$^+_2$, {\it i.e.} the sequential process.
  In spite of the large difference between the non-adiabatic and adiabatic cross sections, the derived reaction rates are concordant with NACRE at the helium burning temperatures.
  Due to the strong influence of 0$^+_2$ in $^{12}$C, astrophysical impact of the direct triple-$\alpha$ process seems to be small.
  It would have been, however, important for theoretical nuclear physicists to understand the off-resonant cross sections, non-adiabatically, and to examine how slow the direct process is at the temperatures relevant to stellar evolution.

  I thank M.~Arnould and Y.~Sakuragi for drawing my attention to the present subject and for their hospitality during my stay at Universit\'e Libre de Bruxelles and Osaka City University.

\providecommand{\href}[2]{#2}\begingroup\raggedright\endgroup

\end{document}